\documentstyle[twocolumn,aps,prl,epsf]{revtex}
\begin{document}
\draft
\title{The Anomalous Magnetic Moment of Quarks}
\author{Pedro J. de A. Bicudo, J. Emilio F. T. Ribeiro and Rui Fernandes}
\address{Departamento de F\'{i}sica and \ \\
Centro de F\'\i sica das Interac\c c\~oes Fundamentais, \\
Edif\'\i cio Ci\^encia, \\
Instituto Superior T\'ecnico, Av. Rovisco Pais, \\
1096 Lisboa, Portugal}
\maketitle

\begin{abstract}
In the case of massless current quarks we find that the breaking of chiral
symmetry usually triggers the generation of an anomalous magnetic moment for
the quarks. We show that the kernel of the Ward identity for the vector
vertex yields an important contribution. We compute the anomalous magnetic
moment in several quark models. The results show that it is hard to escape a
measurable anomalous magnetic moment for the quarks in the case of
spontaneous chiral symmetry breaking.
\end{abstract}

\pacs{}

%\twocolumn[           %
%\widetext             %
%]                     %
%\narrowtext           %
%
%sssssssssssssssssssssssssssssssssssssssssssssssssssssssssssssssssssssssss
%%%%%%%%%%%%%%%%%%%%%%%%%%%%%%%%%%%%%%%%%%%%%%%%%%%%%%%%%%%%%%%%%%%%%

Theoretically, the various hadronic electromagnetic form factors are usually
described in terms of \ pole dominance together with contributions arising
from virtual mesonic exchanges \cite{LitEmCurr}. A third contribution to the
electromagnetic form factors should come from the quark microscopic
interaction itself, in close analogy with QED. It is clear that these three
scenarios should not be independent but just three different aspects of the
same model. This desideratum can be achieved, at least qualitatively, in
terms of a quark field theory displaying spontaneous breaking of chiral
symmetry, ($S\chi SB$).\ In such a description any hadron, when seen from
the trivial vacuum Fock space, appears as a collection of an infinite number
of quark antiquark pairs together with the appropriate valence quarks. It
happens that the contributions of this quark sea can be summarized in terms
of a new set of valence quasiquarks which now carry the information on the
details of the physical vacuum trough a modified propagator \cite{OsArtChiS}%
. In this fashion we recover the simplicity of the constituent quark
picture. It is the role of the Ward identities to ensure charge conservation
throughout this process. And this they do at the expenses of the quark
magnetic moment which, in general, becomes non-zero. As will be shown in
this paper, to maintain throughout the process of $S\chi SB$ a zero
anomalous magnetic moment for the quarks, constitutes the exception rather
than the rule and is just the consequence of particular choices for the
Lagrangian. However, the B.C.S. diagonalization of the Hamiltonian, (mass
gap equation) does not preclude quark pair creation or anihilation processes
to occur. In fact it sets the strength of mesonic contributions for such
physical processes as decay widths and meson-nucleon interactions \cite
{Mes2Vert}, among others. The counterparts of these processes, when seen
from the point of vue of the photon coupling, are precisely pole dominance
and mesonic cloud contributions for the electromagnetic form factors. The
objective of this paper is to set up the general formalism for the
evaluation of electromagnetic form factors in the presence of $S\chi SB$ and
to use it to evaluate the u, d anomalous magnetic moments for various models.

In the Pauli notation, the electromagnetic current up to first order in the
photon momentum $q_{\nu }$ is, 
\begin{eqnarray}
j^{\mu } &=&e\bar{u}\Gamma ^{\mu }u=e\bar{u}\left[ \gamma ^{\mu }+i{\frac{%
\sigma ^{\mu \nu }}{2M}}q_{\nu }a\right] u  \nonumber \\
\mu  &=&\mu _{0}(1+a)\,,\ \mu _{0}={\frac{e\not{h}}{2\;M\;c}}
\label{ElectroCurr}
\end{eqnarray}
where $a$ stands for the anomalous part of the magnetic moment $\mu $ and M
the particle mass. The magnetic moment of ground state hadrons is measured
experimentally. For instance we have for the proton and neutron $a_{p}=1.79$
and $a_{n}=-1.91$. In the constituent quark model for light hadrons we have, 
\begin{eqnarray}
\mu _{p} &=&{\frac{1}{3}}(4\mu _{u}-\mu _{d})=1.852{\mu _{0}}_{p}  \nonumber
\\
\mu _{n} &=&{\frac{1}{3}}(4\mu _{d}-\mu _{u})=-0.972{\mu _{0}}_{p}
\end{eqnarray}
and the quark magnetic moments are nearly proportional to the charges $e_{u}=%
{\frac{2}{3}}e\,,\ e_{d}=-{\frac{1}{3}}e$, which suggests that the
gyromagnetic factor $2(1+a)$ is nearly flavor independent. The quantity
which can be measured is $M/(1+a)$. For quark flavors u and d we have, 
\begin{equation}
M_{u}\simeq (1+a_{u})338\,MeV\ ,\ \ M_{d}\simeq (1+a_{d})322\,MeV
\end{equation}

The constituent quark model can be applied to fit the hadron spectrum, with
a confining interaction, an hyperfine interaction, and a zero point energy, 
\cite{Isgur} The required parameters are of the order of $\alpha
_{s}=0.974\,,\ M_{u}\simeq M_{d}=420\,MeV$ which would suggest a sizeable $a$
of the order of $.15$ to $.3$. It is also clear that we will need $%
a_{d}-a_{u}\simeq .05$ in order to recover the isospin symmetry. 
%%%%%%%%%%%%%%%%%%%%%%%%%%%%%%%%%%%%%%%%%%%%%%%%%%%%%%%%%%%%%%%%%%%%%%%

The Ward Identity, 
\begin{eqnarray}
iq_{\mu }S(p+{q/2})\Gamma ^{\mu } &&S(p-{q/2})=S(p+{q/2})-S(p-{q/2}) 
\nonumber \\
\Leftrightarrow q_{\mu }\Gamma ^{\mu } &&=iS^{-1}(p+{q/2})-iS^{-1}(p-{q/2})
\end{eqnarray}
is obeyed both by the bare vertex $\Gamma _{0}^{\mu }$ and by the
Bethe-Salpeter vertex $\Gamma ^{\mu }$ \cite{Adler}. We will show that in
the limit of small momentum $q$, this identity has the following solution
for the vertex, 
\begin{equation}
\Gamma ^{\mu }(p,q)=i{\frac{\partial }{\partial p_{\mu }}}S^{-1}(p)+q_{\nu }%
{\cal T}^{\nu \mu }(p)+o(q^{2})  \label{vertex kernel}
\end{equation}
where $q_{\nu }{\cal T}^{\nu \mu }(p)$ is defined as the kernel which is not
determined by the Ward identity, 
\begin{equation}
q_{\mu }\left[ q_{\nu }{\cal T}^{\nu \mu }(p)\right] =0\ .
\end{equation}
The Ward identity ensures that charge conservation survives renormalization.
However it does not constrain the kernel, which is a signature of the
renormalization. In particular the kernel contributes to the anomalous
magnetic moment of fermions.

This can clearly be seen in QED where the infrared and ultraviolet
divergences can be removed from the photon propagator, 
\begin{equation}
{\frac{i}{(p^{\prime }-p)^{2}}}\rightarrow {\frac{i}{(p^{\prime
}-p)^{2}-\lambda ^{2}}}-{\frac{i}{(p^{\prime }-p)^{2}-\Lambda ^{2}}}
\end{equation}
The vertex is given by, 
\begin{equation}
\Gamma ^{\mu }=\Gamma _{o}^{\mu }-i\partial ^{\mu }\Sigma +q_{\nu }{\cal T}%
_{o}^{\nu \mu }
\end{equation}
and , up to first order in $\alpha $, the contributions from the self energy
and the kernel to the anomalous magnetic moment are respectively, 
\begin{equation}
\left( -1-2\ln {\frac{\lambda }{M}}\right) {\frac{\alpha }{2\pi }}\ \ ,\ \
\left( 2+2\ln {\frac{\lambda }{M}}\right) {\frac{\alpha }{2\pi }.}
\end{equation}
In the case of actual QED, where $\lambda \rightarrow 0,\Lambda \rightarrow
\infty $, they are both infrared divergent but their sum is finite: $\alpha
/2\pi $.

As for QCD, there has been a considerable effort on how to derive quark
models by integrating out, under various approximations, the gluonic degrees
of freedom. An interesting and promising approach is provided by the
cummulant expansion of the interaction term of the QCD Lagrangian \cite
{Cummulant}. A non-local Nambu Jona-Lasinio type Lagrangian (NJL) is
obtained when we retain only bilocal correlators. Therefore we hold the view
that such quark models are appropriate to study electromagnetic properties
of hadrons, even for light quarks, provided we have small enough photon
momenta and the physics of chiral symmetry breaking is treated correctly.
Therefore, at this stage, rather than focusing on a specific example of NJL
we will study the static electromagnetic properties of a wide class of quark
effective quartic interactions.

In quark models with dynamical $S\chi SB$, the vector vertex $\Gamma ^{\mu }$
is a solution of the Bethe-Salpeter equation,\FL
\begin{eqnarray}
&&\begin{picture}(35,15)(0,0) \put(0,0){$_{p_2}$} \put(0,10){$_{p_1}$}
\put(30,13){$_{q}$} \put(10,0){\line(2,1){10}} \put(10,10){\line(2,-1){10}}
\put(22,5){\oval(4,4)[t]} \put(26,5){\oval(4,4)[b]}
\put(30,5){\oval(4,4)[t]} \end{picture}=\Gamma _{0}^{\mu }+%
\begin{picture}(60,40)(0,0) \put(0,5){$_{p_2}$} \put(0,35){$_{p_1}$}
\put(35,5){$_{p_2'}$} \put(35,35){$_{p_1'}$} \put(2,20){$_{p-p'}$}
\put(60,28){$_{q}$} \put(10,0){\line(1,0){10}} \put(10,40){\line(1,0){10}}
\put(20,0){\line(3,2){30}} \put(20,0){\vector(3,2){20}}
\put(50,20){\line(-3,2){30}} \put(50,20){\vector(-3,2){20}}
\put(52,20){\oval(4,4)[t]} \put(56,20){\oval(4,4)[b]}
\put(60,20){\oval(4,4)[t]} \multiput(20,-2)(0,2){21}{$\cdot$} \end{picture}+%
\begin{picture}(80,40)(0,0) \put(2,10){$_{p_2}$} \put(2,20){$_{p_1}$}
\put(80,23){$_{q}$} \put(30,23){$_{q}$} \put(55,10){$_{p_2'}$}
\put(60,35){$_{p_1'}$} \put(15,10){\line(2,1){10}}
\put(15,20){\line(2,-1){10}} \multiput(25,13)(2,0){8}{$\cdot$}
\put(55,15){\oval(30,30)[]} \put(55,0){\vector(1,0){2}}
\put(55,30){\vector(-1,0){2}} \put(72,15){\oval(4,4)[t]}
\put(76,15){\oval(4,4)[b]} \put(80,15){\oval(4,4)[t]} \end{picture}\nonumber \\
&&p_{1}=p+{\frac{q}{2}},\ p_{2}=p-{\frac{q}{2}},\ p_{1}^{\prime }=p^{\prime
}+{\frac{q}{2}},\ p_{2}^{\prime }=p^{\prime }-{\frac{q}{2}}
\label{vertex diagram}
\end{eqnarray}
where the strong interaction, which is described by a dotted line in the
diagrams is iterated to all orders in the Bethe Salpeter equation. This
equation can be written, 
\begin{eqnarray}
\Gamma ^{\mu }(p,q) &=&\Gamma _{0}^{\mu }-i\int {\frac{d^{4}p^{\prime }}{%
(2\pi )^{4}}}V\left( p^{\prime }-p,p^{\prime }+p,q\right) \Omega
_{a}S(p_{1}^{\prime })  \nonumber \\
&&\Gamma ^{\mu }(p^{\prime },{\frac{q}{2}})S(p_{2}^{\prime })\Omega
_{a}-V\left( q,p^{\prime }+p,-p^{\prime }+p\right) \Omega _{a}  \nonumber \\
&&tr\left\{ S(p_{1}^{\prime })\Gamma ^{\mu }(p^{\prime },q)S(p_{2}^{\prime
})\Omega _{a}\right\}   \label{BSE}
\end{eqnarray}
where the $-1$ factor from the fermion loop was included in the tadpole
term. The momentum dependence of the potential is only assumed to conserve
the total momentum, and in this case it depends on $3$ momenta. The Dirac,
flavor and color structure of the interaction is determined by the $\Omega
_{a}$ matrices. In order to have dynamical $S\chi SB$, we require this
structure to be chiral invariant. Substituting the Ward Identity in the
ladder Bethe Salpeter equation for the vertex we get, \FL
\begin{eqnarray}
&&iS^{-1}(p_{1})-iS^{-1}(p_{2})=iS_{0}^{-1}(p_{1})-iS_{0}^{-1}(p_{2}) 
\nonumber \\
&&-\int {\frac{d^{4}p^{\prime }}{(2\pi )^{4}}}V(p^{\prime }-p,p^{\prime
}+p,q)\Omega _{a}\left[ S(p_{1}^{\prime })-S(p_{2}^{\prime })\right] \Omega
_{a}  \nonumber \\
&&-V\left( q,p^{\prime }+p,-p^{\prime }+p\right) \Omega _{a}tr\left\{ \left[
S(p_{1}^{\prime })-S(p_{2}^{\prime })\right] \Omega _{a}\right\} .
\label{WIBSE}
\end{eqnarray}
For particular cases of the potential $V(p_{1}^{\prime }-p_{1},p_{1}^{\prime
}+p_{2},p_{1}^{\prime }-p_{2}^{\prime })$ we recover the BCS mass gap
equation, \FL
\begin{equation}
{\cal S}^{-1}(p)={\cal S}_{0}^{-1}(p)\ -\ 
\begin{picture}(60,40)(0,0)
\put(25,-8){$_{p'}$} \put(55,0){$_{p}$} \put(17,17){$_{p-p'}$}
\put(50,0){\vector(-1,0){25}} \put(50,0){\line(-1,0){50}}
\put(50.00,0.00){$\cdot$} \put(49.69,3.92){$\cdot$}
\put(48.77,7.72){$\cdot$} \put(47.27,11.84){$\cdot$}
\put(45.22,14.40){$\cdot$} \put(42.67,17.68){$\cdot$}
\put(39.69,20.22){$\cdot$} \put(36.34,22.28){$\cdot$}
\put(32.72,23.78){$\cdot$} \put(28.91,24.70){$\cdot$}
\put(25.00,25.00){$\cdot$} \put(21.08,24.70){$\cdot$}
\put(17.27,23.78){$\cdot$} \put(13.65,22.28){$\cdot$}
\put(10.30,20.22){$\cdot$} \put(7.32,17.68){$\cdot$}
\put(4.77,14.40){$\cdot$} \put(2.72,11.84){$\cdot$} \put(1.22,7.72){$\cdot$}
\put(0.30,3.92){$\cdot$} \put(0.00,0.00){$\cdot$} \end{picture}\ -\ 
\begin{picture}(40,40)(0,0) \put(20,-2){$_{p}$} \put(0,8){$_{0}$}
\put(10,35){$_{p'}$} \put(8,0){\line(1,0){4}}
\multiput(8,-2)(0,2){8}{$\cdot$} \put(10,30){\oval(30,30)[]}
\put(10,45){\vector(-1,0){2}}\end{picture}\label{BCSmassGE} 
\end{equation}
provided that either the rainbow diagram vanishes or, 
\begin{equation}
V(p^{\prime }-p,p^{\prime }+p,q)=V(p^{\prime }-p,p^{\prime }+p,0)\,,
\end{equation}
and that either the tadpole diagram vanishes or, 
\begin{equation}
V(q,p^{\prime }+p,-p^{\prime }+p)=V(0,p^{\prime }+p,-p^{\prime }+p)\,.
\end{equation}
Equation (\ref{BCSmassGE}) can be written, 
\begin{eqnarray}
iS^{-1}(p) &=&iS_{0}^{-1}(p)-\int {\frac{d^{4}p^{\prime }}{(2\pi )^{4}}}%
V(p^{\prime }-p,p^{\prime }+p,0)\Omega _{a}S(p^{\prime })\Omega _{a} 
\nonumber \\
&&-V\left( 0,p^{\prime }+p,-p^{\prime }+p\right) \Omega _{a}tr\left\{
S(p^{\prime })\Omega _{a}\right\} .
\end{eqnarray}
%%%%%%%%%%%%%%%%%%%%%%%%%%%%%%%%%%%%%%%%%%%%%%%%%%%%%%%%%%%%%%%%%%%%%%%%

Now we insert the expression (\ref{vertex kernel}) for $\Gamma ^{\mu }$ in
the Bethe Salpeter equation (\ref{BSE}), in order to find the kernel $q{\cal %
T}$ and expand it up to first order in $q$. The equation for the tensor $%
{\cal T}$, which is antisymmetric, is then, \FL
\begin{eqnarray}
{\cal T}^{\nu \mu } &=&{\cal T}_{0}^{\nu \mu }-i\int {\frac{d^{4}p}{(2\pi
)^{4}}}V\Omega _{a}(1-tr)\left\{ S{\cal T}^{\nu \mu }S\Omega _{a}\right\}  
\nonumber \\
{\cal T}_{0}^{\nu \mu } &=&-{\frac{1}{2}}\int {\frac{d^{4}p}{(2\pi )^{4}}}%
V\Omega _{a}(1-tr)\left\{ {\cal J}^{\nu \mu }\Omega _{a}\right\}   \nonumber
\\
&&{\cal J}^{\nu \mu }={\partial ^{\nu }}(S)\ S^{-1}{\partial ^{\mu }}(S)-{%
\partial ^{\mu }}(S)\ S^{-1}{\partial ^{\nu }}(S)
\end{eqnarray}
This is a self consistent forced linear integral equation. Let us consider a
general quark propagator, solution of the mass gap equation, of the form, 
\begin{equation}
S(p^{\mu })={\frac{iF(p)}{\not{p}-M(p)}}  \label{propagt}
\end{equation}
where $p$=$\sqrt{p^{\mu }p_{\mu }}$. The integrand ${\cal J}^{\nu \mu }$ is
then, 
\begin{eqnarray}
{\cal J}^{\nu \mu } &=&-i{\frac{F}{(p^{2}-M^{2})^{2}}}\Biggl[{\frac{1}{2}}\{%
\not{p},[\gamma ^{\nu },\gamma ^{\mu }]\}+M[\gamma ^{\nu },\gamma ^{\mu }]%
\Biggr.  \nonumber \\
&&\Biggl.-{\frac{\dot{M}}{p}}(p^{\nu }[\not{p},\gamma ^{\mu }]-p^{\mu }[\not%
{p},\gamma ^{\nu }])\Biggr]
\end{eqnarray}
where the dot superscript denotes $d/dp$. In general we find, 
\begin{eqnarray}
{\cal T}^{\nu \mu } &=&t_{1}(p)\{\not{p},[\gamma ^{\nu },\gamma ^{\mu
}]\}+t_{2}(p)\,M[\gamma ^{\nu },\gamma ^{\mu }]  \nonumber \\
&&+t_{3}(p)(p^{\nu }[\not{p},\gamma ^{\mu }]-p^{\mu }[\not{p},\gamma ^{\nu
}])
\end{eqnarray}
Up to $o(q^{2})$ the electromagnetic current of the quark is then, \FL
\begin{eqnarray}
&&j^{\mu }={\frac{e}{F}}\bar{u}\Biggl[\gamma ^{\mu }-{\frac{p^{\mu }}{p}}%
\dot{M}-(\not{p}-M){\frac{p^{\mu }}{p}}{\frac{\dot{F}}{F}}+Fq^{\nu }{\cal T}%
^{\nu \mu }\Biggr]u  \nonumber \\
&=&{\frac{e(1-\dot{M})}{F}}\bar{u}\left[ \gamma ^{\mu }+a\left( i{\frac{%
\sigma ^{\mu \nu }}{2M}}q_{\nu }\right) \right] u\ ,  \nonumber \\
&&a={\frac{\dot{M}+4M^{2}F\left( 2t_{1}+t_{2}\right) }{1-\dot{M}}}
\end{eqnarray}
where the mass shell condition $p=M$ was used together with the Gordon
identities. The anomalous magnetic moment $a$ turns out to be independent of 
$t_{3}$ and $\dot{F}$. However the dependence on $M$ is crucial in models
where $t_{1}$ and $t_{2}$ are finite. In those models $a$ can be thought as
a measure of $S\chi SB$. The quark condensate $\langle \bar{q}q\rangle $ is
also a functional of the dynamically generated mass, 
\begin{equation}
\langle \bar{q}q\rangle \ =\ -\begin{picture}(30,20)(0,0)
\put(20,10){\oval(20,20)[]} \put(20,20){\vector(-1,0){2}} \end{picture}\
=n_{c}\ tr\int {\frac{d^{4}p}{(2\pi )^{4}}}S(p)
\end{equation}
where the trace sums colors with $n_{c}=3$, but the flavor is kept fixed.
Thus, at the onset of the spontaneous $\chi $SB, we will obtain an implicit
relation between $a\,,\ \langle \bar{q}q\rangle $ and the constituent quark
mass, which were simultaneously vanishing before the occurrence of this
phase transition and now become non-zero. 
%%%%%%%%%%%%%%%%%%%%%%%%%%%%%%%%%%%%%%%%%%%%%%%%%%%%%%%%%%%%%%%%%%%

We will now compute $F,\ M,\ a$ and $\langle \bar{q}q\rangle $ in particular
models which are paradigmatic cases of chiral symmetry breaking and comply
with the constraints of the Ward identity.

Model $I$ is the first original NJL model \cite{Nambu1}. The Lagrangian of
model $I$ is, 
\begin{equation}
{\cal L}_{I}=\bar{q}i\not{\partial}q+G\left[ \left( \bar{q}q\right) ^{2}\
-\left( \bar{q}\gamma _{5}q\right) ^{2}\right] 
\end{equation}
where ${\cal L}_{I}$ is specific to the case of 1 flavor, but its results
are similar to the ones of flavor symmetric $U_{A}(n_{f})$ extended NJL
models. The equations will be solved in the momentum representation. As
usual the integrals are done in Euclidean space. A momentum cutoff $\Lambda $
is included in order that the integral in the loop momentum is finite. Since
the cutoff cannot be adscribed to the potential which has to be constant in
momentum space, is must be included in the propagator, 
\begin{equation}
S(p)={\frac{i\ \ F(p)}{\not{p}-M+i\epsilon }}\ ,\ \ F(p)\rightarrow \Theta
_{Euclidian}(\Lambda -p)\,.
\end{equation}
With a constant potential and this momentum cutoff, the loops turn out to be
constant, independent of the external momentum $p$. It is convenient to
evaluate the integrals, 
\begin{eqnarray}
I_{1} &=&-\int {\frac{d^{4}p}{(2\pi )^{4}}}{\frac{iF}{\left(
p^{2}-M^{2}\right) }}  \nonumber \\
&=&{\frac{1}{16\pi ^{2}}}\left[ \Lambda ^{2}-M^{2}\ln \left( 1+{\frac{%
\Lambda ^{2}}{M^{2}}}\right) \right] \,,  \nonumber \\
I_{2} &=&i\int {\frac{d^{4}p}{(2\pi )^{4}}}{\frac{F}{\left(
p^{2}-M^{2}\right) ^{2}}}  \nonumber \\
&=&{\frac{1}{16\pi ^{2}}}\left[ -{\frac{\Lambda ^{2}}{\Lambda ^{2}+M^{2}}}%
+\ln \left( 1+{\frac{\Lambda ^{2}}{M^{2}}}\right) \right] \,,
\end{eqnarray}
where the solid angle $2\pi ^{2}$ is included. The mass gap equation is, 
\begin{eqnarray}
\not{p}-M &=&\not{p}-2G\int {\frac{d^{4}p}{(2\pi )^{4}}}{\frac{iF}{\left(
p^{2}-M^{2}\right) }}\Bigl[(\not{p}+M)\Bigr.  \nonumber \\
&&\Bigl.-\gamma _{5}(\not{p}+M)\gamma _{5}-tr\{\not{p}+M\}\Bigr]
\end{eqnarray}
With the solutions $M=0$ or $1=8\,n_{c}G\,I_{1}(M,\Lambda )$. The parameters 
$\Lambda $ and $G$ are determined once the quark dynamical mass and the
quark condensate are fixed. We now study the kernel in model $I$. Because
the integrals are constant, the antisymmetric tensor ${\cal T}$ is
independent of $p$. Thus ${\cal T}^{\nu \mu }$ has to be of the $t_{2}$
type, proportional to $\left[ \gamma ^{\nu }\,,\,\gamma ^{\mu }\right] $.
Including the structure factors $\Omega _{a}$ we find that the tadpole-like
term vanishes since $\sigma ^{\nu \mu }$ and $\sigma ^{\nu \mu }\gamma _{5}$
have a null trace. In this case of model $I$ the rainbow diagram also
cancels since the structure $1\otimes 1-\gamma _{5}\otimes \gamma _{5}$
projects on the terms with an odd number of Dirac $\gamma $ matrices, of
type $t_{1}$ but $[\gamma ^{\nu }\,,\,\gamma ^{\mu }]$ is even. Thus model $I
$ produces no kernel for the vector vertex and no anomalous magnetic moment
for the quark \cite{DutchandWeise}.

Model $II$ is the second original NJL model \cite{Nambu2}. The Lagrangian
is, 
\begin{equation}
{\cal L}_{II}=\bar{q}i\not{\partial}q+G\left[ \left( \bar{q}q\right) ^{2}\
-\left( \bar{q}\gamma _{5}\vec{\tau}q\right) ^{2}\right] 
\end{equation}
where ${\cal L}_{II}$ is used for 2 flavors $u$ and $d$. It only has an $%
SU(2)_{A}$ symmetry and breaks $U(1)_{A}$ from the onset. Its results are
similar to those of flavor symmetric $SU_{A}(n_{f})$ extended NJL models.
The anzats for the propagator is that of Eq.(\ref{propagt}), and model $II$
only differs from model $I$ in the algebra. The mass gap equation is changed
since $\vec{\tau}\cdot \vec{\tau}=3$ in the fermion line. We get, 
\begin{eqnarray}
\not{p}-M &=&\not{p}-2G\int {\frac{d^{4}p}{(2\pi )^{4}}}{\frac{iF}{\left(
p^{2}-M^{2}\right) }}\Bigl[(\not{p}+M)\Bigr.  \nonumber \\
&&\Bigl.-2{\frac{n_{f}^{2}-1}{n_{f}}}\gamma _{5}(\not{p}+M)\gamma _{5}-tr\{%
\not{p}+M\}\Bigr] \\
\Rightarrow M=0\  &or&\ 1=2\left( 2{\frac{n_{f}^{2}-1}{n_{f}}}%
-1+4n_{f}n_{c}\right) GI_{1}(M,\Lambda )\,.  \nonumber
\end{eqnarray}
were $n_{f}$ and $n_{c}$ stand respectively for the number of flavors and
colors. The rainbow diagram contributes in this case. The tadpole diagram
contribution also changes. The preferred values for the parameters $\Lambda $
and $G$ are $\Lambda =1.65GeV$ and $G=1.23GeV^{-2}$ which yield $M=.33GeV$, $%
\langle \bar{q}\,q\rangle =-(.25GeV)^{3}$ and $f_{\pi }=.09GeV$. As in model 
$I$, the tadpole will not contribute to the antisymmetric tensor which will
be again of the $t_{2}$ type, ${\cal T}^{\nu \mu }=t_{2}\left[ \gamma ^{\nu
},\gamma ^{\mu }\right] $. The first order term is a function of, \FL
\begin{equation}
\int {\frac{d^{4}p}{(2\pi )^{4}}}{\cal J}^{\nu \mu }=-I_{2}\ M\ \left[
\gamma ^{\nu },\gamma ^{\mu }\right] \ .
\end{equation}
In order to evaluate the higher order terms, we calculate, 
\begin{equation}
\int {\frac{d^{4}p}{(2\pi )^{4}}}S\ M\left[ \gamma ^{\nu },\gamma ^{\mu }%
\right] S=-iM^{2}I_{2}\ q_{\nu }\ M\left[ \gamma ^{\nu },\gamma ^{\mu }%
\right] 
\end{equation}
In this case we have two flavors with two different charges $e_{u}={\frac{2}{%
3}}\ ,\ \ e_{d}=-{\frac{1}{3}}$ , and two anomalous magnetic moments $a_{f}$%
, \FL
\begin{eqnarray}
&&e_{u}t_{u}=G\,I_{2}\left[ 0\left( {\frac{e_{u}}{2}}-M^{2}e_{u}t_{u}\right)
+(-2)\left( {\frac{e_{d}}{2}}-M^{2}e_{d}t_{d}\right) \right]   \nonumber \\
&&(u\leftrightarrow d)
\end{eqnarray}
The natural parameter is $2G\,M^{2}\,I_{2}(\Lambda ,M)=0.004$. inverting
this equation we find the solution, \FL
\begin{eqnarray}
a_{u} &\simeq &-2(2GM^{2}I_{2}){\frac{e_{d}}{e_{u}}}=0.004\ ,\ \ \Rightarrow
M_{u}=339\,MeV  \nonumber \\
a_{d} &\simeq &-2(2GM^{2}I_{2}){\frac{e_{u}}{e_{d}}}=0.016\ ,\ \ \Rightarrow
M_{d}=327\,MeV
\end{eqnarray}
Although this effect is small, it has the right sign to correct the $M_{u}$
and $M_{d}$ inversion. If the tadpole term was removed from the mass gap
equation then the $a_{d}-a_{u}$ would be bigger. This is possible for
instance when the potential has a $\vec{\lambda}.\vec{\lambda}$ dependence,
being $\lambda $ the Gell-Mann matrices. 
%%%%%%%%%%%%%%%%%%%%%%%%%%%%%%%%%%%%%%%%%%%%%%%%%%%%%%%%%%%%%%%%%%%%%

Model $III$ is the simplest QCD inspired model. The Lagrangian is, 
\[
{\cal L}_{III}=\bar{q}i\not{\partial}q+{\frac{1}{2}}\bar{q}(x)\gamma
^{\alpha }\frac{\vec{\lambda}}{2}q(x)\cdot \int d^{4}yV(x-y)\bar{q}(y)\gamma
_{\alpha }\frac{\vec{\lambda}}{2}q(y)
\]
In the case of model $III$, the Dirac structure $\gamma ^{\mu }\otimes
\gamma _{\mu }$ is $U_{A}(n_{f})$ chiral invariant. For V(p) we will choose
a color confining square well potential because of its calculational
simplicity. 
\begin{equation}
V(p^{\prime }-p)=-G\ \Theta _{Euclidian}(\Lambda -|p^{\prime }-p|)\ ,
\end{equation}
The mass gap equation is, \FL
\[
{\frac{\not{p}-M}{F}}=\not{p}-{\frac{4i}{3}}\int {\frac{d^{4}p^{\prime }}{%
(2\pi )^{4}}}V(p^{\prime }-p)F{\frac{-2\not{p}^{\prime }+4M}{{p^{\prime }}%
^{2}-M^{2}}}
\]
which includes the color factor of ${\frac{4}{3}}$. We now calculate the
kernel. The first order term for the kernel is a functional of, 
\begin{equation}
\gamma ^{\alpha }{\cal J}^{\nu \mu }\gamma _{\alpha }=iF{\frac{\left\{ \not%
{p},\left[ \gamma ^{\nu },\gamma ^{\mu }\right] \right\} }{\left(
p^{2}-M^{2}\right) ^{2}}}
\end{equation}
The terms with two gamma matrices, of the form $\sigma ^{\mu \nu }$ are now
cancelled by the $\gamma ^{\alpha }\otimes \gamma _{\alpha }$ of the
interaction, and only the $t_{1}$ type term remains. Therefore this model
differs from the previous ones insofar it covers the form factor $t_{1}$.
The self consistent equation for the antisymmetric tensor ${\cal T}$ will
also close, 
\[
\gamma ^{\alpha }S\left\{ \not{p},\left[ \gamma ^{\nu },\gamma ^{\mu }\right]
\right\} S\gamma _{\alpha }=2F^{2}{\frac{p^{2}+M^{2}}{\left(
p^{2}-M^{2}\right) ^{2}}}\left\{ \not{p},\left[ \gamma ^{\nu },\gamma ^{\mu }%
\right] \right\} 
\]
We get, 
\begin{eqnarray}
{\cal T}^{\nu \mu } &=&t\left\{ \not{p},\left[ \gamma ^{\nu },\gamma ^{\mu }%
\right] \right\}   \nonumber \\
t(p) &=&t_{o}-{\frac{8}{3}}i\int {\frac{d^{4}p^{\prime }}{(2\pi )^{4}}}%
V(p^{\prime }-p){\frac{p^{\prime }\cdot p}{p^{2}}}F^{2}(p^{\prime }) 
\nonumber \\
&&{\frac{{p^{\prime }}^{2}+M^{2}}{\left( {p^{\prime }}^{2}-M^{2}\right) ^{2}}%
}t(p^{\prime })\,,  \nonumber \\
t_{o}(p) &=&-{\frac{2}{3}}i\int {\frac{d^{4}p^{\prime }}{(2\pi )^{4}}}%
V(p^{\prime }-p){\frac{p^{\prime }\cdot p}{p^{2}}}{\frac{F(p^{\prime })}{%
\left( {p^{\prime }}^{2}-M^{2}\right) ^{2}}}\,.
\end{eqnarray}
For the euclidean integration it is convenient to evaluate the angular
integrals, 
\begin{eqnarray}
I_{3}( &p^{\prime }&,p)=G\int_{-1}^{+1}dw\theta \left( \Lambda -\sqrt{{%
p^{\prime }}^{2}+p^{2}-2wp^{\prime }p}\right)   \nonumber \\
&=&G(1+I_{6})\theta (1-I_{6})\theta (1+I_{6})+2G\theta (I_{6}-1)>0  \nonumber
\\
I_{4}( &p^{\prime }&,p)=G\int_{-1}^{+1}dw\,w\,\theta \left( \Lambda -\sqrt{{%
p^{\prime }}^{2}+p^{2}-2wp^{\prime }p}\right)   \nonumber \\
&=&G{\frac{1-I_{6}^{2}}{2}}\theta (1-I_{6})\theta (1+I_{6})>0  \nonumber \\
I_{5} &=&{\frac{\Lambda ^{2}-{p^{\prime }}^{2}-p^{2}}{2p^{\prime }p}}
\end{eqnarray}
and the nonlinear integral mass gap equation for $F$ and $M$, the integral
for $t_{o}$, the linear integral equation for $T$, and the integral for $%
\langle \bar{q}q\rangle $ can be solved simultaneously, \FL
\begin{eqnarray}
F(p) &=&\left[ 1+\int_{0}^{\infty }dp^{\prime }{\frac{I_{5}(p^{\prime },p)}{%
6\pi ^{2}p}}{\frac{{p^{\prime }}^{4}}{{p^{\prime }}^{2}+M^{2}(p^{\prime })}}%
F(p^{\prime })\right] ^{-1}  \nonumber \\
M(p) &=&F(p)\int_{0}^{\infty }dp^{\prime }{\frac{I_{4}(p^{\prime },p)}{3\pi
^{2}}}{\frac{{p^{\prime }}^{3}F(p^{\prime })}{{p^{\prime }}%
^{2}+M^{2}(p^{\prime })}}M(p^{\prime })  \nonumber \\
t_{o}(p) &=&\int_{0}^{\infty }dp^{\prime }{\frac{I_{5}(p^{\prime },p)}{24\pi
^{2}p}}{\frac{{p^{\prime }}^{4}F(p^{\prime })}{\left[ {p^{\prime }}%
^{2}+M^{2}(p^{\prime })\right] ^{2}}}  \nonumber \\
t(p) &=&t_{o}-\int dp^{\prime }{\frac{I_{5}(p^{\prime },p)}{6\pi ^{2}p}}{%
\frac{{p^{\prime }}^{4}F(p^{\prime })^{2}\left[ {p^{\prime }}%
^{2}-M^{2}(p^{\prime })\right] }{\left[ {p^{\prime }}^{2}+M^{2}(p^{\prime })%
\right] ^{2}}}t(p^{\prime })  \nonumber \\
\langle \bar{q}q\rangle  &=&-\int_{0}^{\infty }dp{\frac{3}{2\pi ^{2}}}{\frac{%
p^{3}F(p)M(p)}{p^{2}+M^{2}(p)}}
\end{eqnarray}
The mass term has a trivial solution $M=0$ and another solution which breaks
spontaneously chiral symmetry. A dimensional simplification occurs if we
work in units of $\Lambda =1$. In this case the only parameter is $G$ which
is now adimensional. We find a critical value $G_{c}=132$ above which chiral
symmetry occurs. In Fig. 1 we depict the values of $M$, $\langle \bar{q}%
\,q\rangle $ and $a$. We solve the integral equations numerically for $F\
,\,M$ and $t$ with the Gauss iterative method and using the Gauss
integration \cite{Liu}. We find that at $p^{2}=-1$ these functions decrease
by a factor of just $.9\rightarrow .7$. Since we cannot continue
analytically the numerical solution we use the approximation of nearly
constant $F,\,M$ and $t$ and compute the mass and the anomalous magnetic
moment for $p=0$. The literature prefers a $<\bar{q}q>=-(0.25\,GeV)^{3}$. A
dynamical quark mass $M=0.33\,GeV$ would correspond to $G=245\Lambda ^{-2}$, 
$\Lambda =0.74\,GeV$ and $a=0.15$ . If we now consider a $M=(1+a)\,0.33\,GeV$
then the lowest possible condensate is $<\bar{q}q>=-(0.28\,GeV)^{3}$ which
corresponds to $G=300\Lambda ^{-2}$, $\Lambda =0.69\,GeV$, $M=0.42\,GeV$ and 
$a=0.28$ , see Fig. 1. 
%%%%%%%%%%%%%%%%%%%%%%%%%%%%%%%%%%%%%%%%%%%%%%%%%%%%%%%%%%%%%%%%%%%%%%%%

NJL models $I$ and $II$ are the simplest models with chiral symmetry
breaking. In the NJL model $I$ the anomalous magnetic moment $a$ vanishes.
In the model $II$ the $U(1)$ breaking interaction yields a too small $a$,
which nevertheless provides an example of an isospin dependence for $a$ and,
therefore, contributes to the $u\ -\ d$ mass inversion. The reason for the
smallness of the anomalous magnetic moment stems from the presence of
tadpole contributions and were not for this contribution and we would have
obtained a much larger $a$. This is precisely the case of model $III$ where
a larger $a$ is derived, compatible with the nonrelativistic constituent
quark models. We also find that M, a, $<\bar{q}q>$, are functions of $%
(G-G_{c})$ with critical exponents which are respectively $1$, $2$ and $1$.
The present work constitutes a first step on a more elaborate model unifying
hadronic spectroscopy (including decay widths) and the electromagnetic form
factors which have been shown to be consistent with the simple quark
constituent picture precisely because of $S\chi SB$.

%%%%%%%%%%%%%%FIGURA%%%%%%%%%%%%%%%%%%%%%%%%%
\begin{figure}
\begin{picture}(400,200)(0,0)
\put(-50,-30){\epsffile{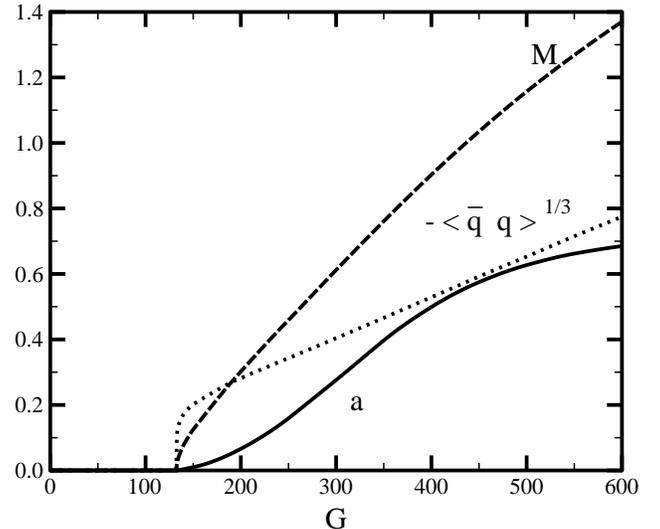}}\end{picture}
%\input{Epsf.sty} 
%\begin{figure}[hbt]
%\mbox{\epsfxsize = 8 cm \epsfysize =11 cm \epsffile{fig1.eps}}
\caption{Functions of the quark dynamical mass M, anomalous magnetic moment
a and quark condensate as functions of the adimensional coupling G for model
III }
\end{figure}


\begin{references}
\bibitem{LitEmCurr}  H. Ito,Phys. Rev C {\bf 52}, R1750, (1195); H. Forkel,
M. Nielsen, X Jin, T. Cohen, Phys Rev C {\bf 50}, 3108, (1994).

\bibitem{OsArtChiS}  A. Le Yaouanc, L.Oliver, O. Pene and J-C. Raynal, Phys
Rev D {\bf 29}, 1233 (1984); A. Le Yaouanc, L.Oliver, S. Ono, O. Pene and
J-C. Raynal, {\it ibid} {\bf 31}, 137 (1985);P.Bicudo and J. Ribeiro, Phys
Rev D {\bf 42}, 1611 (1990).

\bibitem{Mes2Vert}  P.Bicudo and J.Ribeiro Phys Rev D {\bf 42}, 1635 (1990);
P.Bicudo and J.Ribeiro Phys Rev C {\bf 55}, 834 (1997).

\bibitem{Isgur}  N. Isgur, G. Karl, Phys. Rev. D {\bf 18}, 4178

\bibitem{Adler}  S. Adler and A. C. Davis, Nucl. Phys. B {\bf 224}, 469
(1984).

\bibitem{Cummulant}H. G. Dosh, Phys Lett. {\bf B190} 
177 (1987); H. G. Dosh and U. Marquard, Nuc Phys. {\bf A560} 333
(1993); N. Brambilla and A. Vairo Phys. Lett {\bf B407} 167 (1997); Yu. A.
Simonov hep-ph/9712248.

\bibitem{Nambu1}  Y. Nambu, G. Jona-Lasinio, Phys. Rev. {\bf 122}, 345 (1961)

\bibitem{DutchandWeise}  U. Vogl, M. Lutz, S. Klimt and W. Weise, Nucl.
Phys. A {\bf 516}, 469 (1990); J. Singh, Phys. Rev. D {\bf 31}, 1097 (1985).

\bibitem{Nambu2}  Y. Nambu, G. Jona-Lasinio, Phys. Rev. {\bf 124}, 246
(1961); V. Bernard, Phys. Rev. D {\bf 34}, 1601 (1986).

\bibitem{Liu}  Y. Dai, Z. Huang and D. Liu, Phys. Rev. D {\bf 43}, 1717
(1991).
\end{references}
\end{document}